\documentclass[twocolumn,letter,latterpaper]{jpsj3}
\usepackage{txfonts}
\usepackage{bm}

\title{Multiple Superconducting Phases and Unusual Enhancement of the Upper Critical Field in UTe$_2$}

\author{
Dai~Aoki$^{1,2}$\thanks{E-mail: aoki@imr.tohoku.ac.jp}, 
Fuminori~Honda$^1$,
Georg~Knebel$^2$,
Daniel~Braithwaite$^2$, 
Ai~Nakamura$^1$,
DeXin~Li$^1$,
Yoshiya~Homma$^1$,
Yusei~Shimizu$^1$,
Yoshiki~J.~Sato$^1$,
Jean-Pascal~Brison$^2$, and
Jacques~Flouquet$^2$
}

\inst{%
$^1$IMR, Tohoku University, Oarai, Ibaraki, 311-1313, Japan\\
$^2$Univ. Grenoble Alpes, CEA, IRIG, PHELIQS, F-38000 Grenoble, France\\
}

\abst{%
We performed AC calorimetry and magnetoresistance measurements under pressure for $H\parallel a$-axis (easy-magnetization axis) in the novel heavy-fermion superconductor UTe$_2$.
Thanks to the thermodynamic information, multiple superconducting phases have been revealed under pressure and magnetic field.
The ($H,T$) phase diagram of superconductivity under pressure displays an abrupt increase of the upper critical field ($H_{\rm c2}$) at low temperature and in the high field region, and a strong convex curvature of $H_{\rm c2}$ at high temperature. 
This behavior of $H_{\rm c2}$ and the multiple superconducting phases require a state for the superconducting order parameter more complex than a spin-triplet equal spin pairing.
Above the superconducting critical pressure, $P_{\rm c}$, we find strong indications that the possible magnetic order is closer to antiferromagnetism than to ferromagnetism.
}

\begin{document}
\maketitle
The recently discovered heavy-fermion superconductivity (SC) in UTe$_2$ attracts much attention~\cite{Ran19,Aok19_UTe2},
because spin-triplet SC is most likely realized in this system, 
lying at the proximity of ferromagnetic (FM) order.
SC coexisting microscopically with long-range FM order is already well studied~\cite{Aok12_JPSJ_review,Aok19} in three uranium compounds, UGe$_2$~\cite{Sax00}, URhGe~\cite{Aok01} and UCoGe~\cite{Huy07}.
Because of the strong internal field due to the FM moment, a spin-triplet state with equal-spin pairing (ESP) is favored. 
In this case, SC can survive even under strong internal exchange field.
Furthermore SC can be even reinforced at high magnetic field ($H$), pressure ($P$) and uniaxial stress by tuning the FM fluctuations.
When the field is applied along the intermediate hard-magnetization axis ($b$-axis) in URhGe and UCoGe,
the FM Curie temperature is suppressed, and the FM fluctuations are remarkably enhanced.
Then field-reentrant (-reinforced) SC is observed at high fields, which highly exceeds the so-called Pauli limit~\cite{Lev05,Aok09_UCoGe}. 

A quite similar situation might be also realized in UTe$_2$.
UTe$_2$ is a paramagnet with a body-centered orthorhombic crystal structure (space group: $Immm$, {\#}71, $D_{2h}^{25}$).
The large Sommerfeld coefficient, $\gamma\sim 120\,{\rm mJ\,K^{-2}mol^{-1}}$, indicates  strong correlations in the electronic states.~\cite{Ike06_UTe2} 
The magnetization curves show a relatively large anisotropy between the easy-magnetization axis ($a$-axis) and the hard-magnetization axes ($b$ and $c$-axes).
The $b$-axis is the hardest magnetization axis at low temperatures; 
the magnetization curve shows a sharp 1st order metamagnetic transition at $H_{\rm m}=35\,{\rm T}$,
where the effective mass is strongly enhanced.~\cite{Miy19,Kna19,Ima19}
The value of $H_{\rm m}$ is well scaled by $T_{\chi_{\rm max}}$ ($\sim35\,{\rm K}$), at which a broad maximum of susceptibility $\chi$ is observed at low field.

The SC properties of UTe$_2$ are spectacular. 
The SC transition occurs at $T_{\rm sc}=1.6\,{\rm K}$ with the large specific heat jump.
The residual $\gamma$-value amounts to $\sim 40\,{\%}$ against the normal state $\gamma$-value for the best quality sample,~\cite{Aok20_SCES} and the entropy balance is not satisfied, assuming a constant $\gamma$-value extrapolated from the normal state above $T_{\rm sc}$. 
Strong electronic correlations are dominant in this system, which is also indirectly confirmed by the failure of direct LDA band structure calculations;
they predict an Kondo insulating ground state~\cite{Har20,Aok19_UTe2}, and large enough Coulomb repulsion ($U$) is required to turn it into a metallic state \cite{Ish19,Xu19}.
Hence, the upper critical field is large and anisotropic, 
$H_{\rm c2}=6.5\,{\rm T}$ and $11\,{\rm T}$ for the $a$ and $c$-axes, respectively.
For the $b$-axis, field-reentrant SC appears at high fields and survives up to $H_{\rm m}=35\,{\rm T}$. 
At $H_{\rm m}$, SC is abruptly suppressed.
Since the values of $H_{\rm c2}$ for all the field directions highly exceed the Pauli-limit,
a spin-triplet state seems established.
Furthermore, when the field direction is tilted from $b$ to $c$-axis (hard- to hard-axis) by $\sim 30\, {\rm deg}$, SC reappears above $H_{\rm m}\sim 40\,{\rm T}$.~\cite{Ran19_HighField}
From a microscopic point of view, the development of FM fluctuations is detected by $\mu$SR~\cite{Sun19} and NMR~\cite{Tok19} experiments. 
No long-range magnetic order was detected down to very low temperatures by $\mu$SR and magnetization~\cite{Pau20} measurements. 
The very small decrease of the NMR Knight shift below $T_{\rm sc}$ supports a spin-triplet scenario~\cite{Nak19}. 
A point node gap structure was proposed by the specific heat, thermal conductivity and penetration depth measurements~\cite{Met19,Kit20}. 
Furthermore, topological SC is suggested by experiments and theory.~\cite{Bae19,Jia19,Ish19}

Applying  pressure also revealed surprising behaviors; 
$T_{\rm sc}$ initially decreases slightly and then increases above $0.3\,{\rm GPa}$ up to $\sim 3\,{\rm K}$ at $1.2\,{\rm GPa}$.~\cite{Bra19,Ran19_pressure}
Further increasing pressure, SC is suppressed at $P_{\rm c}\sim 1.5\,{\rm GPa}$,
and a new ordered state which probably corresponds to the magnetic order (MO) appears at $T_{\rm m}=3\,{\rm K}$. 
In addition, remarkably, multiple SC phases are detected at zero field by the AC calorimetry measurements above $0.2\,{\rm GPa}$~\cite{Bra19}.

In order to study the multiple SC phases of UTe$_2$ in details, 
we have performed AC calorimetry measurements as a thermodynamic probe to
determine ($H,T$) phase diagrams.  Magnetic field has been applied along the easy-magnetization axis ($a$-axis) at different pressures.
An unusual enhancement of $H_{\rm c2}$ at low temperature and in high field regions was found.
Multiple SC phases are found inside the SC domain under magnetic field and pressure.
Our experiments give here unique insights, which could not be observed in magnetoresistance measurements restricted to the $H_{\rm c2}$ boundary between normal and SC phases.~\cite{Kne20}

Single crystal growth and experimental technique of AC calorimetry and magnetoresistance measurements are described in Ref.\citen{Suppl}.

\begin{figure}[tbh]
\begin{center}
\includegraphics[width=\hsize]{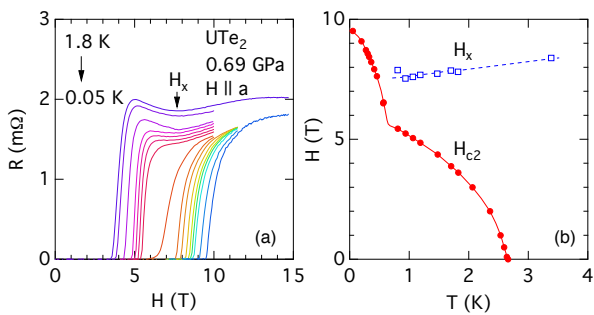}
\end{center}
\caption{(Color online) (a) Magnetoresistance for $H\parallel a$-axis at $0.69\,{\rm GPa}$ in UTe$_2$
at different temperatures, 1.8, 1.7, 1.5, 1.2, 1.1, 0.94, 0.81, 0.56, 0.45, 0.41, 0.36, 0.32, 0.30, 0.27, 0.19 and 0.05 K. The electrical current was applied along $a$-axis. (b) ($H,T$) phase diagram for $H\parallel a$-axis at $0.69\,{\rm GPa}$ determined from the magnetoresistance measurements in UTe$_2$.}
\label{fig:MR}
\end{figure}

Figure~\ref{fig:MR}(a) shows the field dependence of the magnetoresistance for $H\parallel a$-axis at $0.69\,{\rm GPa}$ at different  temperatures. 
The electrical current was applied along the $a$-axis.
At the lowest temperature, $0.05\,{\rm K}$, $H_{\rm c2}$ defined by the zero resistivity is enhanced up to $9.5\,{\rm T}$ ($H_{\rm c2}\sim 6.5\,{\rm T}$ at ambient pressure).
At higher temperatures, a broad minimum appears at $H_x\sim 7.5\,{\rm T}$ in the normal state.
This anomaly may correspond to that detected at ambient pressure ($\sim 6\mbox{--}10\,{\rm T}$) by magnetoresistance~\cite{Kna19}, magnetization~\cite{Miy19} and thermopower~\cite{Niu19}, which is most likely due to a Lifshitz transition with Fermi surface reconstruction.

Figure~\ref{fig:MR}(b) shows the ($H,T$) phase diagram at $0.69\,{\rm GPa}$ determined from the results of magnetoresistance. 
The SC transition temperature is $2.65\,{\rm K}$ at zero field and $H_{\rm c2}$ is $9.5\,{\rm T}$ at $0.05\,{\rm K}$.
The temperature dependence of $H_{\rm c2}$ plotted by solid circles is quite unusual. 
Firstly, $H_{\rm c2}$ suddenly increases with a kink at $0.6\,{\rm K}$ upon cooling.
Secondly, the $H_{\rm c2}$ curve at high temperature shows a strong convex (downward) curvature with a large initial slope at $T_{\rm sc}$ ($-dH_{\rm c2}/dT = 8.1\,{\rm T/K}$). 
This curvature could be an indication of Pauli paramagnetic limitation.
Note that a weak convex curvature is also observed in specific heat measurements at ambient pressure~\cite{Kit20}, and a similar behavior under pressure has already been observed on magnetoresistance measurements~\cite{Kne20}.
The broad anomaly denoted by $H_x$ is almost constant with slight increase with temperature and disappears above $3.5\,{\rm K}$.

\begin{figure}[tbh]
\begin{center}
\includegraphics[width= \hsize]{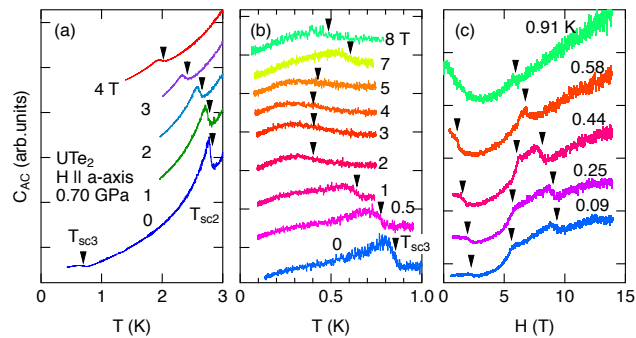}
\end{center}
\caption{(Color online) (a) Temperature dependence of the AC calorimetry at different fields at $0.70\,{\rm GPa}$ for $H\parallel a$-axis in UTe$_2$. (b) Temperature dependence of the AC calorimetry obtained by the phase analysis at different fields. (c) Field dependence of the AC calorimetry at different temperatures. The data are vertically shifted for clarity.}
\label{fig:C_AC1}
\end{figure}

Figure~\ref{fig:C_AC1} shows the results of AC calorimetry measurements at $0.70\,{\rm GPa}$.
In Fig.~\ref{fig:C_AC1}(a), the temperature dependence of AC calorimetry ($C_{\rm AC}$) at zero field shows double transitions at $T_{\rm sc2}=2.8\,{\rm K}$ and $T_{\rm sc3}=0.85\,{\rm K}$.
The results are in good agreement with the previous report~\cite{Bra19}, 
although $T_{\rm sc2}$ is slightly higher here.
With field, $T_{\rm sc2}$ shifts to  lower temperature.
On the other hand, $T_{\rm sc3}$ slightly decreases with field and then increases again, as shown in Fig.~\ref{fig:C_AC1}(b).
Here $C_{\rm AC}$ was obtained from the phase analysis, that is, $C_{\rm AC} \propto 1/\tan(\phi)$, where $\phi$ is the phase calculated from the real and imaginary components of the signal in the lock-in amplifier. 
At $5\,{\rm T}$ the signal is rather small, but the anomaly is still visible. 
At higher field, $7\,{\rm T}$, the anomaly becomes sharper again.
The field scan of AC calorimetry at different temperatures shows several anomalies, as shown in Fig.~\ref{fig:C_AC1}(c).
At $0.09\,{\rm K}$, three anomalies are detected at $9.3$, $5.9$ and $2.2\,{\rm T}$. 
The first anomaly corresponds to $H_{\rm c2}$ detected by the magnetoresistance as described above.
Note that the field scan of AC calorimetry was measured at the constant frequency,
thus the increase or decrease of the signal at anomaly is arbitrary because of the field variation of thermal conductivity, though the anomaly clearly displays the phase boundary.

\begin{figure}[tbh]
\begin{center}
\includegraphics[width= 0.9\hsize]{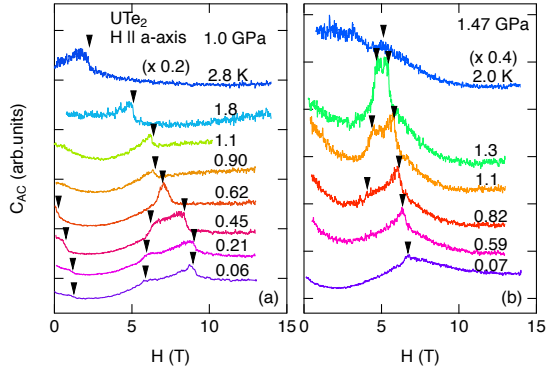}
\end{center}
\caption{(Color online) Field dependence of AC calorimetry at $1.0\,{\rm GPa}$ (a) and $1.47\,{\rm GPa}$ (b) for $H\parallel a$-axis in UTe$_2$. The data at $2.8\,{\rm K}$ ($2.0\,{\rm K}$) at $1.0\,{\rm GPa}$ ($1.47\,{\rm GPa}$) are scaled by factor, 0.2 (0.4).
The data are vertically shifted for clarity.}
\label{fig:C_AC_Hscan}
\end{figure}

At $1.0\,{\rm GPa}$, three transitions are detected at $0.99$, $5.8$ and $9.1\,{\rm T}$ in the field scan at the lowest temperature, $0.06\,{\rm K}$ as shown in Fig.~\ref{fig:C_AC_Hscan}(a). 
The lower field transition is suppressed with increasing temperature, and the higher two transitions merge at $0.62\,{\rm K}$.
Note that the transitions at $1.1$ and $1.8,{\rm K}$ lead to marked anomalies.
In the temperature scan at constant fields, $2$--$5\,{\rm T}$, no anomaly was detected below $1\,{\rm K}$, in contrast to the case at $0.7\,{\rm GPa}$.

At $1.47\,{\rm GPa}$ which is presumably just below the SC critical pressure $P_{\rm c}$, a single transition is detected at $6.7\,{\rm T}$ at $0.07\,{\rm K}$ in the field scan, as shown in Fig.~\ref{fig:C_AC_Hscan}(b).
The lower field transition starts to appear above $0.8\,{\rm K}$ (at $4\,{\rm T}$) and the two transitions merge at $2\,{\rm K}$. 
The temperature scan at low fields shows no anomalies. 
The detected anomalies in the field scan seems a survival of SC, as shown later in the ($H,T$) phase diagram.

At high pressures above $P_{\rm c}$ ($\sim 1.5\,{\rm GPa}$), SC is suppressed and a new ordered phase, most likely magnetic, appears at $T_{\rm m}$.
The details are described in Ref.\citen{Suppl}.

\begin{fullfigure}[tbh]
\begin{center}
\includegraphics[width= 0.8\hsize]{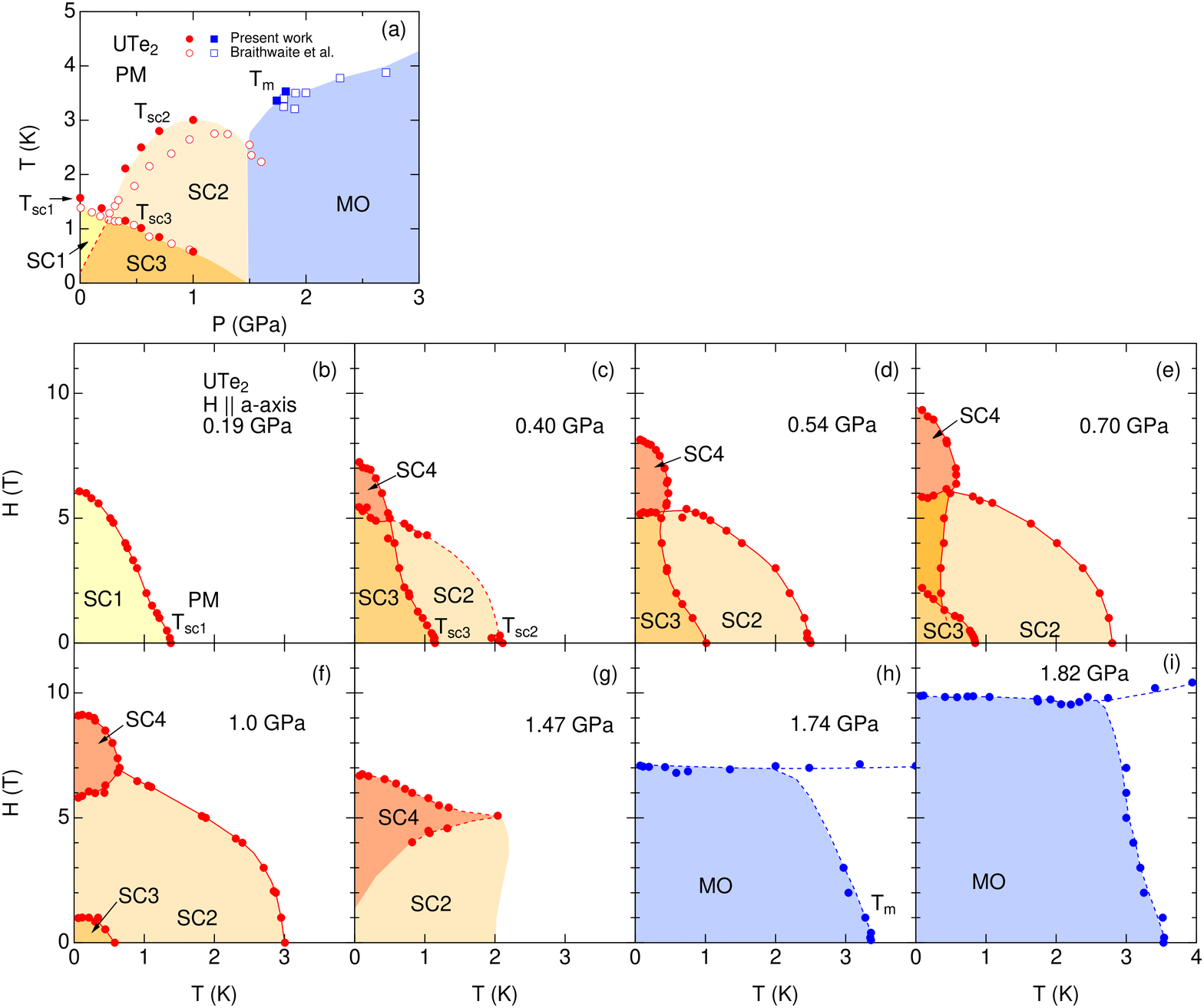}
\end{center}
\caption{(Color online) (a) ($T,P$) phase diagram at zero field in UTe$_2$. The solid circles/squares are the present work. The open circles/squares are cited from Ref.\citen{Bra19}. (b)-(i) ($H,T$) phase diagrams for $H\parallel a$-axis at different pressures. The lines are guides to the eyes.}
\label{fig:HT_phase}
\end{fullfigure}
Figure~\ref{fig:HT_phase}(a) shows the ($T,P$) phase diagram at zero field.
The results are consistent with the previous ones~\cite{Bra19}. 
With increasing pressure, $T_{\rm sc1}$ initially decreases, then splits above $0.2\,{\rm GPa}$.
The lower transition temperature $T_{\rm sc3}$ decreases linearly with pressure and extrapolates to zero at $P_{\rm c}\sim 1.5\,{\rm GPa}$.
The higher transition temperature $T_{\rm sc2}$ increases under pressure and has a maximum near $1\,{\rm GPa}$.
In the normal state, $T_{\rm m}$ detected above $P_{\rm c}\sim 1.5\,{\rm GPa}$  is also in good agreement with the previous results.

In Fig.~\ref{fig:HT_phase}(b)-(i), we show all the ($H,T$) phase diagrams at different pressures.
It is obvious that multiple SC phases exist at and above $0.4\,{\rm GPa}$.
Thus, UTe$_{2}$ clearly belongs to the rare case of (spin-triplet) superconductors with multiple SC states, like UPt$_3$~\cite{Bru90}, Th-doped UBe$_{13}$~\cite{Shi17} , and of course, superfluid $^3$He~\cite{Leg75}.

The ($H,T$) phase diagram at $0.19\,{\rm GPa}$ is similar to that at ambient pressure. 
At $0.40\,{\rm GPa}$, in zero field, two transitions $T_{\rm sc2}$ and $T_{\rm sc3}$ appear.
The lower temperature $T_{\rm sc3}$ shows a significant concave (upward) curvature with ``S''-shape.
On the other hand, the higher temperature $T_{\rm sc2}$ shows a convex (downward) curvature.
At $0.54\,{\rm GPa}$, the ``S''-shaped behavior of $T_{\rm sc3}$ and the convex (downward) shape of $T_{\rm sc2}$ appear clearly.
The phase boundary between normal and superconducting state at $0.7\,{\rm GPa}$ nearly coincides with that at $0.69\,{\rm GPa}$ detected by magnetoresistance measurements shown in Fig.~\ref{fig:MR}(b).
The ``S''-shape of $T_{\rm sc3}$ shows a vertical increase from 2 to $6\,{\rm T}$.
$T_{\rm sc3}$ seems to split further (at $0.4\,{\rm K}$ and $1.3\,{\rm T}$), indicative of the evolution of the phase diagram at higher pressure. 
At $1\,{\rm GPa}$, the ``S''-shaped behavior is disconnected, and two decoupled phases appear at high field and low field regions.
The strong convex (downward) curvature for $T_{\rm sc2}$ is very remarkable at pressures between $0.4$ and $1\,{\rm GPa}$.
At $1.47\,{\rm GPa}$ just below the critical pressure $P_{\rm c}$ of SC, 
no anomaly is detected in the temperature scan at zero field,
while a clear double transition in the field scan associated with SC is detected in the limited temperature range between $0.8$ and $1.4\,{\rm K}$.
At this pressure, $T_{\rm sc2}$ is rapidly suppressed as the first order transition in the ($T,P$) phase diagram, as shown in Fig.~\ref{fig:HT_phase}(a)
Therefore it is quite difficult to detect $T_{\rm sc2}$ by a temperature scan.
Note that the magnetoresistance at $1.4\,{\rm GPa}$ for $H\parallel a$-axis reveals SC at $1.9\,{\rm K}$ at zero field; nearly vertical initial slopes of $H_{\rm c2}$ are detected in the ($H,T$) phase diagram close to $P_{\rm c}$.~\cite{Kne20}

At higher pressure above $P_{\rm c}$, anomalies are clearly detected again at zero field, 
and the ($H,T$) phase diagrams are shown in Fig~\ref{fig:HT_phase}(h)(i).
The anomaly at $T_{\rm m}$ shifts to  lower temperature with field, and it  collapses at $H_{\rm c}\sim 7$ and $10\,{\rm T}$
for $1.74$ and $1.82\,{\rm GPa}$, respectively.
For other field directions ($H\parallel b$ and $c$-axes)~\cite{Braithwaite,Knafo}, $T_{\rm m}$ also decreases and is suppressed under magnetic field. 
Thus the magnetic phase above $P_{\rm c}$ may be not a FM order.
If the order is FM, $T_{\rm m}$ should increase rapidly and change into a crossover when the field is applied along the easy-magnetization axis. 
In the Ising FM case, $T_{\rm m}$ would decrease for the field along hard-magnetization axis as observed in URhGe and UCoGe.
The decrease of $T_{\rm m}$ for all the field directions in UTe$_2$ is suggestive of an antiferromagnetic (AFM) order.
The phase diagram shown in Fig.~\ref{fig:HT_phase}(h)(i) is similar to the AFM one known in the heavy fermion systems.
$H_{\rm c}$ survives far above $T_{\rm m}$ up to $5$ and $7\,{\rm K}$ for $1.74$ and $1.82\,{\rm GPa}$, respectively.
At $1.82\,{\rm GPa}$, $H_{\rm c}$ slightly increases from $10$ to $11\,{\rm T}$ above $T_{\rm m}$.
The shape of the anomaly at $5\,{\rm K}$ is clearly different from those below $T_{\rm m}$, as shown in Ref.~\citen{Suppl}.
These results imply that the first order transition occurs at low temperatures, and that a crossover occurs at high temperature above $T_{\rm m}$.
Similar magnetic phase diagrams are known in  heavy fermion antiferromagnets, such as UPd$_2$Al$_3$~\cite{Sug93} and CeRh$_2$Si$_2$~\cite{Kna17}, where a crossover field is observed at high temperatures above $T_{\rm N}$.

A very striking feature of our results is SC4 marked by an abrupt increase of $H_{\rm c2}$ observed at low temperatures for pressures between $0.4$ and $1\,{\rm GPa}$. As $H_{\rm c2}$ shows signs of Pauli limitation, an explanation could be that it marks the appearence of a Fulde-Ferrell-Larkin-Ovchinnikov (FFLO) state. Such a state has been recently revealed in the ``two dimensionnal'' (2D) iron-based superconductors, KFe$_2$As$_2$~\cite{Cho17}, and FeSe~\cite{Kas20}.
In the FFLO state, the pair-breaking due to the Pauli paramagnetic effect is reduced, 
because Cooper pairs between the two Zeeman split Fermi surfaces are formed with a finite $\bm{q}$ for their center of mass.
In UTe$_2$, a FFLO scenario could also be relevant, even for its spin-triplet state, precisely because it could enhance the observed paramagnetic limit (arising from a finite $\langle S_Z=0\rangle$ component of the superconducting order parameter for that field direction). However, it is very unlikely that it plays any role in the observed abrupt increase of $H_{\rm c2}$: first, because UTe$_2$ is a 3D superconductor, so the enhancement of $H_{\rm c2}$ should not be larger than $6\%$; secondly, because the FFLO state never produces such a sharp kink, only an inflection point (see e.g. Refs.~\citen{Cho17,Kas20} ); and last, because it is clear from our thermodynamic measurements, that the increase emerges as a prolongation of the lower SC transition $T_{\rm sc3}$, existing already at zero field, where no paramagnetic limitation exists. 

Besides the identification of the symmetries of the different phases, 
and the microscopic origin of such a complex SC phase diagram, 
a major question is the observation of a strong convex (downward) curvature of $H_{\rm c2}(T)$ for $H\parallel a$-axis.
The main point, by analogy with the ferromagnetic superconductors, is that it is difficult to imagine that the most favored spin-triplet state has paramagnetic limitation along the field directions with the largest susceptibility~\cite{Ike06_UTe2} and spin fluctuations~\cite{Tok19,Sun19}.

In order to solve this paradox, a phenomenological model has been proposed in Refs.~\citen{Kit20,Mac20}, where the superconducting order parameter ($\bm{d}$-vector) would have indeed no intrinsic paramagnetic limitation along the a-axis, but where $T_{\rm sc}$ would be initially field-enhanced in this direction due to the development of a strong magnetization.
This may work for the weak effects detected at zero pressure, but certainly cannot be applied to explain the very strong curvature observed at and above $0.4\,{\rm GPa}$. 
From our pressure measurement, it appears that the transition line at $T_{\rm sc3}$ is reminiscent of $H_{\rm c2}(T)$ at  $0.19\,{\rm GPa}$, where we do not observe any paramagnetic limitations for $H \parallel a$-axis
(keeping in mind that $T_{\rm sc3}(H)$ is a transition between two superconducting phases, not an $H_{\rm c2}$ line).

Therefore the strong paramagnetic limitation observed on $H_{\rm c2}$ ($T_{\rm sc2}(H)$) should not be compared to the observation at zero pressure \cite{Kit20}.
At the opposite, it suggests that $T_{\rm sc2}$ emerges from an order parameter different from that at zero pressure, 
originating from another, more efficient, pressure-induced pairing mechanism.
This would be coherent with the observation that the pressure induced magnetic order above $P_{\rm c}$ is not simple FM, so that magnetic fluctuations of a different type develop at the proximity of $P_{\rm c}$.
Most likely, owing to the very large $H_{\rm c2}$ observed along the $c$-axis in magnetoresistance measurements, with no sign of paramagnetic limitation in this directions, it could be that the order parameter is still the spin-triplet, but with a complex non-unitary $\bm{d}$-vector having no component along the $c$-axis. 
The phase boundary between SC3 and SC2 would be associated with a rotation of the $\bm{d}$-vector, and SC4 would mark the restoration of the low pressure dominant order parameter (at least up to $0.9\,{\rm GPa}$, the ``easy-magnetization'' axis remains the $a$-axis.)

In UTe$_2$, as in 4$f$ or 5$f$ heavy-fermion systems, pressure modifies the local properties of U.
This is well studied in Ce compounds, where the Kondo effect, crystal field and volume drastically change with pressure.~\cite{Flo09}
As reported in the recent paper,~\cite{Kne20}
the $P$ variation of the magnetocrystalline anisotropy is a key player in the SC and PM properties.
A fascinating new ingredient in this system is the stabilization of a metallic state at $P=0$; 
pressure could induce a critical interplay between hybridization and Coulomb repulsion, 
leading to drastic changes with reinforcement of sharp electronic anomalies of the density of states.

In recent reports with field scans along the $b$-axis,~\cite{Kne20,Lin20} the link between the pressure collapse of the metamagnetic transition at $H_{\rm m}$, and the $H_{\rm c2}$ curves has been established.
Contrary to the case of FM-SC, the order parameter cannot be restricted to the spin-triplet state with ESP.
Thus, as shown here, field can select different order parameters with well defined spin and orbital components,
as it occurs for the clean case of superfluid $^3$He with its A, A$_1$ and B phases.
A striking point is the strong convex curvature of $H_{\rm c2}$ clearly observed here above $0.54\,{\rm GPa}$.
In our thermodynamic measurements, multiple SC phases of UTe$_2$ have been revealed in the ($H,T,P$) domain.
Furthermore, even if the precise nature of the MO phase above $P_{\rm c}$ remains to be solved, it appears much closer to an AF phase than a FM one. 
So, this opens the possibility that FM fluctuations collapse at $P_{\rm c}$ at the development of AF order. 
The relation with a possible valence shift by pressure from U$^{3+}$ to U$^{4+}$ should be further explored, as well as the pressure evolution of the band structure. 
The ($T,P$) phase diagram of UTe$_2$ remind us of the case of Tm chalcogenides, where band structure, valence and magnetic ordered states are known to be in strong interplay.~\cite{Ant02,Lin98,Las83,Der06}

\section*{Acknowledgements}
We thank K. Machida, Y. Yanase, V. Mineev, S. Fujimoto, W. Knafo, K. Ishida and Y. Tokunaga, 
for fruitful discussion.
This work was supported by ERC starting grant (NewHeavyFermion), KAKENHI (JP15H05884, JP15H05882, JP15K21732, JP16H04006, JP15H05745, JP19H00646), and ICC-IMR.


\section*{Supplemental Material}

High quality single crystals of UTe$_2$ were grown using chemical vapor transport (CVT) method with Iodine as the transport agent.~\cite{Aok20_SCES} 
The magnetoresistance under pressure was measured by the four-probe DC method using a MP35N piston cylinder cell.
The AC calorimetry measurements under pressure were performed in another MP35N piston cylinder cell, using a Au/Fe-Au thermocouple and thin Au wires as the thermometer and the heater, respectively.
The thermocouple and the heater were directly spot-welded on the small sample (approximately $0.5\times 0.3\times 0.1\,{\rm mm}^3$ in dimension), and the signal was detected by a lock-in amplifier with typical frequency of about $115\,{\rm Hz}$.
Daphne 7373 oil was used as  pressure transmitting medium for both piston cylinder cells.
The pressure was determined by the SC transition temperature of Pb.
Both magnetoresistance and AC calorimetry measurements were done in a top-loading dilution refrigerator at low temperatures down to $0.05\,{\rm K}$ and at high fields up to $15\,{\rm T}$.

\begin{figure}[tbh]
\begin{center}
\includegraphics[width= 0.8\hsize]{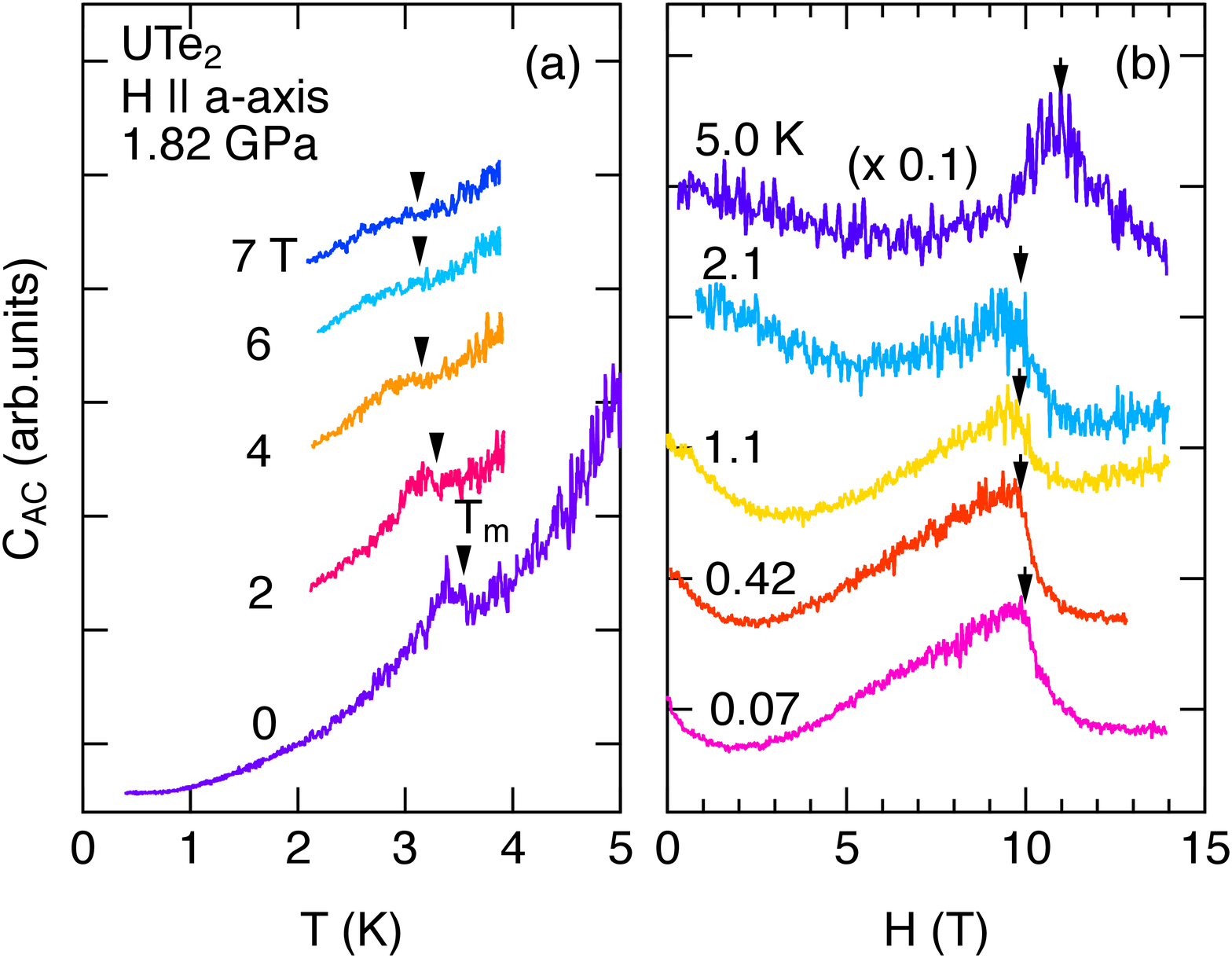}
\end{center}
\caption{(Color online) (a) Temperature dependence of the AC calorimetry at different fields for $H\parallel a$-axis at $1.82\,{\rm GPa}$ in UTe$_2$. (b) Field dependence of the AC calorimetry at different temperatures. The data at $5\,{\rm K}$ was scaled by factor 0.1 for comparison. The data are vertically shifted for clarity.}
\label{fig:C_AC2}
\end{figure}
Figure~\ref{fig:C_AC2}(a) shows the temperature dependence of the AC calorimetry at $1.82\,{\rm GPa}$.
At zero field, a jump corresponding to the magnetic order is observed at $T_{\rm m}=3.5\,{\rm K}$.
The jump becomes smaller under magnetic fields and shifts slightly to lower temperatures.
As shown in Fig.~\ref{fig:C_AC2}(b), the field scan reveals a clear anomaly at $10\,{\rm T}$ at the lowest temperature. 
The anomaly survives even at high temperatures above $T_{\rm m}$ up to $7,{\rm K}$, but it becomes progressively a broad maximum, indicating a crossover at high temperatures rather than a phase transition.

\end{document}